\documentclass{article}

\usepackage{arxiv}

\usepackage[utf8]{inputenc} % allow utf-8 input
\usepackage[T1]{fontenc}    % use 8-bit T1 fonts
\usepackage{hyperref}       % hyperlinks
\usepackage{url}            % simple URL typesetting
\usepackage{booktabs}       % professional-quality tables
\usepackage{amsfonts}       % blackboard math symbols
\usepackage{nicefrac}       % compact symbols for 1/2, etc.
\usepackage{microtype}      % microtypography
\usepackage{graphicx}
\usepackage{natbib}
\usepackage{doi}
\usepackage{array}

\title{Decolonisation, Global Data Law, and Indigenous Data Sovereignty}

\date{April, 30, 2022} 			

\author{ \href{https://orcid.org/0000-0002-6710-4950}{\includegraphics[scale=0.06]{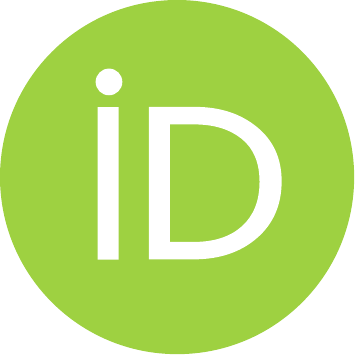}\hspace{1mm}Jennafer Shae Roberts} \\
	Accel AI Institute\\
	San Francisco, CA \\
	\texttt{jennafershae@accel.ai} \\
	\And
	\href{https://orcid.org/0000-0001-8115-8851}{\includegraphics[scale=0.06]{orcid.pdf}\hspace{1mm}Laura N Montoya} \\
	Accel AI Institute\\
	San Francisco, CA  \\
	\texttt{laura@accel.ai} \\
}

\renewcommand{\shorttitle}{Decolonisation, Global Data Law, and Indigenous Data Sovereignty}

\hypersetup{
pdftitle={A template for the arxiv style},
pdfsubject={q-bio.NC, q-bio.QM},
pdfauthor={Jennafer S. Roberts, Laura N. Montoya},
pdfkeywords={Indigenous data sovereignty, data law, decolonisation, neo-colonialism, data colonialism, digital colonialism.},
}

\begin{document}
\maketitle

\begin{abstract}
	This research examines the impact of digital neo-colonialism on the Global South, and encourages the development of legal and economic incentives to protect Indigenous cultures globally. Data governance is discussed in an evolutionary context, while focusing on data sharing and data mining. Case studies which exemplify the need to steer global data law towards protecting the earth, while addressing issues of data access, privacy, rights, and colonialism in the global South are explored. The case studies highlight connections to indigenous people's rights, in regards to the protection of environmental ecosystems, thus establishing how data law can serve the earth from an autochthonous lens. This framework examines histories shaped by colonialism and suggests how data governance could be used to create healthier balances of power.
\end{abstract}

% keywords can be removed

\section{Introduction}
Policymakers must understand that in efforts to decolonise, they must take care to prevent reinforcing colonial and neo-colonial practices when dealing with data and data law on a global scale. Decolonisation cannot occur by building on top of old oppressive colonial systems. Rather, decolonisation exists in the cracks, and in the cracks, revolution is made. This paper examines some of those cracks, and widens them by defining terms that give us the language to talk about decolonisation, in an effort to instigate positive change. Situated from an Indigenous perspective, case studies from around the globe are presented to decentralise on the Global North agenda, with a focus on taking care of the natural world. Ethics of data and data law are central tenets within this paper. First we will lay a foundation for this discussion by defining data colonialism and digital colonialism as modern forms of colonisation and neo-colonisation. Next, Indigenous data sovereignty (ID-SOV)\citep{Kukuta2016IndigenousSovereignty} will be explored, as it is imperative to include the voices and perspectives of Indigenous peoples throughout the world when discussing decolonisation. The next section discusses global data law and highlights movements which could stand to be of aid in the decolonial turn. Following, the colonial practices of data mining are discussed. Then an effort is made to connect ID-SOV, data law, and the need to protect the natural environment. In the final section, the Global South is complexified and case studies from the Inuit are presented to further discuss decolonising, starting with research, and some controversial effects of digital colonialism. 

This research is largely influenced by ID-SOV and also expands on research by Nick Couldry and Ulises A. Mejias\citep{Couldry2019DataSubject}\citep{Couldry2021TheHeading} on data colonialism and the decolonial turn in the digital sphere and the impacts of it. We also turn to the conclusions made by many other researchers throughout our discussion which relate to ID-SOV and decolonisation. This paper highlights various movements and organisations in positions to inform decolonisation for data law and to inform and enliven ID-SOV. We cover case studies from across the world, including Africa\citep{Abebe2021NarrativesAfrica}, South America,\cite{Rojas-Paez2021NarrativesSetting} and Northern Canada\citep{CunsoloWillox2013StorytellingWisdom}, to give a broad view of ID-SOV, digital and data colonialism, global data law and decolonisation. This is an overview that is far from exhaustive, but gives fresh perspectives by stressing the need to prioritise the voices of Indigenous peoples as equally sovereign leadership.

\section{Understanding Ethics amid Neo-Colonialism, Digital Colonialism and Data Colonialism
}
Colonialism is a deeply rooted world system of power and control that plays out in ways that have become normal, yet are far from anything that would be considered ethical. In our modern world which is so reliant on technology, colonialism and neo-colonialism within data and the digital realm remain a fundamental problem. In 1965, Kwame Nkrumah, the first President of Ghana and founding member of the Non-Alignment Movement, defined neo-colonialism as when the State outwardly has all the markers of international sovereignty, however, foreign capital is used for exploitation and imperialist powers have control. In this way, neo-colonial investment increases rather than decreases the gap between wealthy and impoverished countries.\citep{Tiger1966Neo-Colonialism.Imperialism} This is a more subtle yet dangerous form of colonialism that still exists today, as it affects everyone around the world, and is being amplified in the digital sphere.

There is a strong separation between the dominant powers, and the people and communities from which they profit. This is often perpetuated by seeing the Global North as separate from the Global South. Stefania Milan and Emiliano Treré presented a plural definition of the South(s) as place and proxy going beyond geopolitical denomination and embracing a multiplicity of interpretations, creating space for anywhere that people “. . . suffer discrimination and/or enact resistance to injustice and oppression and fight for better life conditions against the impending data capitalism.”\citep{Milan2019BigUniversalism} We use the terminology of Global North and Global South broadly, but this review references examples which are not specific to this aforementioned framework. One such instance regarding digital colonialism affecting Inuit communities in Northern Canada is key to our exploration. 

There are two ways neo-colonialism is being discussed in sociotechnical language: digital colonialism and data colonialism. These are parallel terms and may be considered one in the same, however we will focus here on how they have been described independently.

\label{}

\subsection{Digital Colonialism}
When digital technology is used for social, economic, and political domination over other nations or territories, it is considered digital colonialism. Dominant powers have ownership over digital infrastructure and knowledge, which perpetuates a state of dependency within the hierarchy, situating Big Tech firms at the top and hosting an extremely unequal division of labour, which further defines digital colonialism.\citep{Kwet2021DigitalEmpire.}

\subsection{Data Colonialism}

Data colonialism addresses Big Data in the context of the predatory practices of colonialism. Capitalism depends on the data from the Global South, which represents a new type of appropriation attached to current infrastructures of connection.\citep{Couldry2019DataSubject} We see a pulley system of interdependence. However, the concentration of power is clear. 

Data colonialism is sometimes seen as a subset of digital colonialism, as data colonialism concerns the abstraction of life into bits and bytes, whereas digital colonialism encompasses data but also the infrastructure and hardware that make up the digital world and the connection to the internet.\citep{Mouton2021DigitalCity} 

\subsection{Decolonisation and Capitalism} 
We cannot address colonialism without also addressing capitalism. Colonialism came first, and historical colonialism, with its violence and brutality, paved the way for capitalism.\citep{Couldry2021TheHeading} In order to decolonize, we need to fully overhaul the systems of capitalism and consumerism. We cannot add on or amend laws or regulations to govern data and the digital world in an attempt to decolonize without inadvertently reproducing colonialism. We need a full system change, and it is going to take a lot of work. 

We are at the dawn of a new stage of capitalism, following the path laid out by data colonialism, just as historical colonialism paved the way for industrial capitalism. We cannot yet imagine what this will look like, but we know that what lies at its core is the appropriation of human life through the misuse of data.\citep{Couldry2019DataSubject} 

Not only is this a problem because it creates global inequality, capitalism notably threatens the natural environment. Its structural imperative is based on an insatiable appetite for growth and profit, causing overconsumption of Earth’s material resources, not to mention overheating the planet.\cite{Kwet2021DigitalEmpire.} Mining cobalt in the Congo has detrimental effects not just on the earth, but on people's lives, utilizing harsh child labor.\cite{Lawson2021TheMining} The Congo is where we get over 50\% of the world's cobalt, an essential raw mineral found in cell phones, computers and electric vehicles, as well as in lithium batteries, in which we will see an increase in demand with the rise of renewable energy systems.\citep{BanzaLubabaNkulu2018SustainabilityCongo} Furthermore, not only is data mining causing harm to people and the environment in how it is being collected but also how it is being stored long-term. Data centres\citep{BetterBuildingsInitiative2021DataCenters} alone account for 2\% of human carbon emissions, rivalling that of the airline industry.\cite{AirTransportActionGroup2022ATAG} There are plans and efforts to lower emissions\citep{InternationalEnergyAgency2022DataNetworks} from data centres, which needs to be done across industries, alongside efforts to address the underlying issues of dependence due to capitalism and consumerism. 

\subsection{Decolonial Thinking and Examples}
 “What decolonial thinking in particular can help us grasp is that colonialism — whether in its historic or new form — can only be opposed effectively if it is attacked at its core: the underlying rationality that enables continuous appropriation to seem natural, necessary and somehow an enhancement of, not a violence to, human development.’’\citep{Couldry2019DataSubject} 

In this paper, we will provide many examples that explore various corners of the world and the impact of digital and data colonialism in different ways, including data mining, and case studies in the African Indigenous context.\citep{Abebe2021NarrativesAfrica} Within data mining, we will discuss how or even if data mining is different from data sharing, as well as contextualise data mining alongside resource mining from the Earth. 

Further examples include the impact of internet usage in Indigenous  communities such as the Inuit,\citep{Young2019TheColonialism} where local knowledge is waning due to the influence of digital colonialism, as well as in Mexico and Colombia,\cite{Walter2021INDIGENOUSPOLICY} where we further examine environmental concerns related to Indigenous rights and data governance. In order to have a truly ethical AI, there needs to be a major shift in societal ethics, and decolonisation of data and the digital world is the primary focus for this work. 

Now that we have introduced the concepts of colonisation, neo-colonisation, digital colonialism and data colonialism, it is time to consider Indigenous Data Sovereignty and the principles upon which it relies. We will situate Indigeneity throughout history and understand the importance of focusing on Indigenous rights today in decolonising for global governance of data.

\section{In Consideration of Indigenous Data Sovereignty
}
“Indigenous Peoples have always been ‘data warriors’. Our ancient traditions recorded and protected information and knowledge through art, carving, song, chants and other practices.” 

Before computers, even before the development of the written language, knowledge and data were continuously passed down from generation to generation by Indigenous Peoples, and yet Indigenous Data Sovereignty (ID-SOV) is a relatively new concept, first being published in 2016.\citep{Kukuta2016IndigenousSovereignty} This section will review ID-SOV and the CARE principles of Indigenous data governance in an effort to move towards decolonizing data. 

ID-SOV has been defined as “. . . the right of Indigenous Peoples to own, control, access and possess data that derive from them, and which pertain to their members, knowledge systems, customs or territories.”\citep{Kukuta2016IndigenousSovereignty} 

To state ID-SOV as a human right is one thing; however to see it carried out, we must detangle from a long history of manipulation of data on Indigenous  peoples, who were historically demonised and dehumanised to justify settler colonialism. Today, when neo-colonialism is rife, we see how this narrative has transformed, yet is now presented by victimising Indigenous peoples. This narrative needs to change in order to empower Indigenous peoples. According to the Global Indigenous Data Alliance (GIDA), building strategic relationships with global bodies and mechanisms is necessary to promote ID-SOV and governance internationally by providing a visible, collective approach.\citep{Kukuta2016IndigenousSovereignty} 

Even in recent times, sensitive COVID-19 data has been mined and reused without consent from Indigenous Americans by the media, researchers and non-governmental organisations. This has been carried out under the assumption that publicising Indigenous communities’ sensitive data would be helpful, but actually causing unintentional harm in the process.\citep{RDACOVID-19WorkingGroup2020RDASharing} Settler colonialists thought that they were ‘helping’ too, via ethnic cleansing and direct violence. Where neo-colonialism is not inherently violent, it is extremely dangerous ,\cite{Couldry2021TheHeading} and tracing the histories can help us understand how to move towards decolonizing data for the benefit of all. 

\subsection{Decolonising Data Via Self-Determination}
Data and data analytics have become increasingly important and interdependent in many ways in the digital age. Governments are heavily reliant on data for policy and decision making. As has been the case in much of our history, the unwilling targets of policy interventions are disproportionately Indigenous Peoples, whose enduring aspirations for self-determination over their own knowledge, information systems, institutions and resources are consistently undermined by larger consolidated governments. Data is mined from Indigenous Peoples, as well as their lands and cultures without their input or permission on how their data is collected, used or applied.\citep{Walter2021INDIGENOUSPOLICY} However, as will be explored throughout this paper, ID-SOV and governments have been striving to change this and pull away from such a victimising narrative. 

\subsection{What does it mean to be Indigenous?}
To have the conversation about ID-SOV, let us first discuss the challenge in defining what it means to be Indigenous. According to The UN Declaration on the Rights of Indigenous Peoples (UNDRIP), indigeneity has to do with first colonial contact, which can be quite difficult to determine in countries where colonists were not settlers. The term ‘tribes’, although useful, is problematic In its colonial origins. However, ‘Indigenous’ can encompass a wide range of ethnically diverse peoples, including tribes, such as the hill tribes in the Mekong River area of Southeast Asia.\citep{Scott2009TheAsia} A common element of Indigenous Peoples is a strong desire to maintain autonomy, while also resisting marginalisation and discriminations, which are often discussed in the mainstream with a victimising narrative.\citep{Chung2019INDIGENOUSREGION} 

While trying to decolonise, we must stress that the term Indigenous was a separation created by colonists used to determine who was human, and who was less than human.\citep{Scott2009TheAsia} The fact that we still function from this foundation is inherently detrimental. In working towards decolonizing data in modern times, the place to start is with Indigenous Data Sovereignty. If we centre on the rights of those who have been most marginalised by colonialism, methods by which we can continue the process of decolonization will become progressively clearer.

\subsection{Indigenous Data Concerns}
There are a wide array of data concerns from Indigenous  groups, such as those in the Mekong area of Southeast Asia, referred to as Indigenous ethnic minorities (IEM). Many contradictions arise that result in security risks, and the impact of sharing IEM data could be both positive and negative in unanticipated ways. A balance of freedoms is required - transparency must be weighed in conjunction with individual security.\citep{Chung2019INDIGENOUSREGION} 

Within this contradiction lies a major difficulty: how to have accessible and transparent data, while also ensuring the right to privacy for the subjects of that data. This presents a much deeper issue - data does not promote change automatically nor does it address issues of marginalization, colonialism or discrimination. Additionally, there is little to no consideration given to combatting imbalances of power in negotiations and consultations led by major world governments.\citep{Chung2019INDIGENOUSREGION}

Open Data movements are concerning for ID-SOV networks due to the lack of protection for Indigenous Peoples. There is an increased push for greater data sharing, which can be seen in the widely-accepted FAIR principles (Findable, Accessible, Interoperable, Reusable), however this created tensions in regards to how Indigenous Peoples’ data is protected, shared and used. In order to encourage data collectors and users to be more aligned with Indigenous worldviews, the CARE Principles are a framework to use for consideration of appropriate data use.\citep{Kukutai2020IndigenousSovereignty}  

\subsection{CARE Principles for Indigenous Data Governance}
While the FAIR principles are data-centric and ignore the impact on ethical and socially responsible data usage, including power differentials and historic conditions considering the acquisition and usage of data, the CARE principles centre on the well-being of Indigenous  Peoples and their data, and can be implemented alongside the FAIR Principles throughout data lifecycles to ensure collective benefit.\citep{RDACOVID-19WorkingGroup2020RDASharing}

\begin{table}
	\caption{CARE Principles}
	\centering
	\begin{tabular}{ | m{9em} | m{12cm}| } 
		\toprule
		\cmidrule(r){1-2}
		Name     & Description \\
		\midrule
		Collective Benefit & Data ecosystems shall be designed and function in ways that enable Indigenous Peoples to derive benefit from the data.\\
		\hline
		Authority to Control & Indigenous Peoples’ rights and interests in their own data must be recognised and their authority to control such data must be empowered. Indigenous data governance enables Indigenous Peoples and governing bodies to determine how they, as well as their lands, territories, resources, knowledge and geographical indicators, are represented and identified within data. \\
		\hline
		Responsibility &  Those working with Indigenous data have a responsibility to share how those data are used to support Indigenous Peoples’ self determination and collective benefit. Accountability requires meaningful and openly available evidence of these efforts and the benefits accruing to Indigenous Peoples. \\
		\hline
		Ethics &  Indigenous Peoples’ rights and well-being should be the primary concern at all stages of the data life cycle and across the data ecosystem. \\
		\bottomrule
	\end{tabular}
	\label{tab:table}
\end{table}

If these principles can be integrated into systems of open data, it could truly turn towards decolonizing data. However, they need to be more than just principles. If we centre on the CARE principles and ID-SOV for data governance on a global scale, perhaps we can steer away from harmful colonial data mining and towards a more balanced relationship with data. Before we dive deeper into data mining, the next section addresses data law more directly and highlights movements which could be in positions to do the difficult work of decolonising data governance. 

\section{Global Data Law and Decolonisation
}
Anywhere on Earth with an internet connection, AI systems can be accessed from the cloud, and teams from different countries can work together to develop AI models, which rely on many datasets from across the planet along with cutting edge machine learning techniques.\citep{Engler2022Regulation} 

This global nature of AI can allow for many collaborative projects, however, it can also perpetuate ongoing marginalisation and exploitation of the people behind the data that machine learning relies on. Data protection laws vary per country, and in the U.S., per state, making it difficult for businesses who are trying to maintain ethical principles,\citep{Barber2021NavigatingSociety} while simultaneously creating targets, (mostly in the under-protected ‘Global South’) for large corporations to mine data freely and gain power,\citep{Couldry2019DataSubject} keeping inequality alive and well. What can we do about this? Is the development of a Global Data Law the answer?

As the US and the EU work to attempt to align on data protection,\citep{Engler2022Regulation} there is a call for a re-enlivening of the Non-Aligned Movement (NAM) in the digital sphere.\citep{Mejias2020ToMovement} \citep{Reddy2021Movement} The NAM is an anti-colonial/anti-imperialist movement currently consisting of 120 countries, which are not aligned with any major world powers. It was founded in 1961 to oppose military blocs during the Cold War, and has not yet been adopted for opposing data colonialism, although some say that it should.\cite{Reddy2021Movement} This is intriguing because we are working with a plethora of cultures which have variant value systems, and aligning with Big Government and corporate powers is not beneficial across the board. In fact, it can be quite harmful. When approaching the idea of global data law, we must prioritise the rights of people who are the most marginalised. “What we need is a Non-Aligned Technologies Movement (NATM), an alliance not necessarily of nations, but of multiple actors already working towards the same goals, coming together to declare their non-alignment with the US and China.”\citep{Mejias2020ToMovement} Ulises Ali Mejias calls for NATM to be more society - and community - driven than generating state powers, which makes it all the more viable for the current situation.

The concept of decolonisation is central to this approach to global data law. If we are not careful, neo-colonialism will lead to yet another stage of capitalism that is fueled by unethical misuse of data. That is why it is essential to involve Indigenous and marginalised peoples, not at the margins but at the centre of debates on global standards for law and data, or else efforts to decolonise will only serve to reinforce colonialism.\citep{Couldry2019DataSubject} 

However, colonialism and neo-colonialism are strong systems which feed imperial powers, whether they be governmental or corporate. The root of capitalism - prioritising exponential financial growth - stands in the way of creating lasting, tangible change that will not reproduce harmful and outdated systems in new ways. 

\subsection{What is Global Data Law?}
Global data law is an area of great tension, for it is necessary to regulate and protect sensitive data from around the world, whether people live in the EU under protection of the General Data Protection Regulation (GDPR), in other countries with strong data protection laws, or in areas with less data protection. Creating a “one-size fits all” system of global data laws will not work in our diverse world with varying tolerances for oversight and surveillance. 
 
The GDPR is being used as a model for data protection, but it only protects the privacy of EU citizens, no matter where in the world the data is used.\citep{Reddy2021Movement} Compliance with the GDPR is mandatory, requiring complete transformation in the way organisations collect, process, store, share and wipe personal data securely. Otherwise there are exorbitant fines in the tens of millions of dollars designed to penalise data misuse.\citep{DLAPiperGlobalLawFirm2022EUFirm} Several other countries have developed specific data protection laws in the last few years, scattered across the world, for example in Canada and Brazil. However the protections vary greatly. The details of all of the countries with data laws and without is beyond the scope of this paper, and new data laws are being developed as we write.
 
\subsection{UN Recommendations}
We can also consider the UN recommendations, such as to move away from data ownership and towards data stewardship for data collectors, meanwhile protecting privacy and ensuring self-determination of peoples’ own data.\citep{TheUN2022UNStrategy} The UN stresses that there is a need to protect basic rights of peoples’ data not being used or sold without permission and in ways that could cause undue harm, with respect to what this means across cultures. 

The UN Roadmap for Digital Cooperation highlights: 
\begin{itemize}
	\item global digital cooperation
	\item digital trust and security
	\item digital human rights
	\item human and institutional capacity building
	\item an inclusive digital economy and society\citep{TheUN2022UNStrategy}
\end{itemize}
To date, what we are seeing in global data law is that governments are developing their own systems unilaterally, making compliance complicated. For example, in the first years of the GDPR, thousands of online newspapers in the US simply decided to block users from the EU rather than face compliance risks.\citep{Freuler2020TheMovement} \citep{South2018MoreEffect}  Those in the EU were not able to access information previously available to them. Businesses that rely on customers from the EU had to trade off the risk of the compliance liability to the loss of income from their European clientele.
 
Global data law is indeed complex, and to further complicate matters, next we will briefly dive deeper into the Digital Non-Alignment movement. 

\subsection{Situating the Digital Non-Aligned Movement}
The original Non-Aligned Movement (NAM) was formed by leaders of many countries, mainly from the Global South, which sought a political space to counter central powers through coordinated solidarity. A central tenet was to exercise strategic autonomy, resisting control from the US and the USSR during the Cold War.\citep{South2018MoreEffect} \citep{Freuler2020TheMovement} In today’s digital age, there is a call to recentre on the NAM in the digital realms to protect against not only government powers, but Big Tech as well.\citep{Freuler2020TheMovement}

A Non-Aligned Technologies movement (NATM) would empower civil societies across the globe to act in concert to meet their shared objectives while putting pressure on their respective governments to change the way they deal with Big Tech. The primary goal of NATM would be to transition from technologies that are detrimental to society to technologies that are actually in the interest of a healthier and more equilateral society.\citep{Mejias2020ToMovement} 

“NAM must once again come together to ensure the free flow of technology and data, while simultaneously guaranteeing protection to the sovereign interests of nations.”\citep{Reddy2021Movement} 

This point is incredibly valid, and countries represented within the NAM deserve to have a voice in this discussion about global data law. The US, China, the EU, and other wealthy nations should not be solely responsible for the global regulation of open data and sovereignty. However, sovereignty is not just for states, which is why Indigenous Data Sovereignty (ID-SOV) should also be used for guiding global data law towards decolonisation: if we are going to decolonise, we must centre on the rights of those who have been the most harmed by colonisation.

\subsection{Turning to Indigenous Data Sovereignty to Inform Global Data Law}

Indigenous Peoples’ focus on self-determination is continuously burdened with the implications of data collected and used against them. The UN Declaration of the Rights of Indigenous Peoples (UNDRIP) states that the authority to control Indigenous cultural heritage (i.e. Indigenous data: their languages, knowledge, practices, technologies, natural resources and territories) should belong to Indigenous communities.\citep{Carroll2020TheGovernance} It proves to be extremely difficult to break free from colonial and neo-colonial structures of power imbalances, however; this is exactly what must be the focus in order to decolonise data practices and data law. 

We are up against a long history of extraction and exploitation of value through data, representing a new form of resource appropriation that could be compared to the historical colonial land-grab, where not only land and resources but human bodies and labour were seized, often very violently. The lack of outright violence in today’s data colonialism does not negate its danger. “The absence of physical violence in today’s data colonialism merely confirms the multiplicity of means by which dispossession can, as before, unfold.”\citep{Couldry2019DataSubject}  

Contemporary data relations are laced with unquestionable racism,\citep{Couldry2019DataSubject} along with intersectional discrimination against all those considered marginalised, which is why we turn next to a report that addresses these issues directly, with a focus on surveillance and criminalization via data in the US. 

\subsection{Highlighting Technologies for Liberation}

The report titled Technologies for Liberation\citep{Neves2020TechnologiesLiberation} arose from the need to better understand the disproportionate impact of surveillance and criminalization of  Queer, Trans, Two-Spirit, Black, Indigenous, and People of Color (QT2SBIPOC) communities and provide a resource for these communities to push back and protect themselves at all levels, from the state-endorsed to the corporate-led.\citep{Neves2020TechnologiesLiberation} Technologies for Liberation aims to decolonise at a grassroots, community level, focusing on organisers and movement technologists which visualise demilitarised, community-driven technologies that support movements of liberation. This is transformative justice at work, centering on safety and shifting power to communities.\citep{Neves2020TechnologiesLiberation} This is the bottom-up influence on data protection that we need to be turning towards to inform global data law that won’t leave people on the margins. 

\subsection{Listening to inform Global Data Law}

There are endless areas that need consideration when discussing global data law and decolonisation. We have already touched on a few areas and highlighted movements such as the digital NAM, ID-SOV, and Technologies for Liberation, after introducing the GDPR and the UN recommendations. We have discussed potential avenues for solutions, including organisational principles and values which are key to this discussion. There are large power imbalances that need to be addressed and deeply rooted systems that need to be reimagined. Therefore, we must start with listening to the voices who have been the most silenced, and allow them to start the process of guiding everyone involved to do the right thing.

“The time has come for us to develop a set of basic principles on which countries can agree so that consumers worldwide are protected and businesses know what is required of them in any geography.”\citep{Barber2021NavigatingSociety} 

In the following section, data mining is problematised as a colonial practice in an effort to understand how to move forward and decolonise data governance.

\section{The Ethics of Data Sharing and Data Mining}

The collection and use of data which relies on peoples’ production and sharing of personal and sensitive information has a certain ‘creep factor.’ When it is used in ways which fail to consider the people behind the data, the creep factor increases dramatically. 
 
For example, the media, researchers, and non-governmental organisations continue to access and reuse sensitive data without consent from Indigenous governing bodies. One case in point has occured recently within the COVID-19 pandemic where tribal data in the United States was released by government entities without permission or knowledge of the communities themselves. There is an effort to address gaps in data and data invisibility of Indigenous peoples in America, however this can result in unintentional harm while ignoring Indigenous sovereign rights.\citep{RDACOVID-19WorkingGroup2020RDASharing} 
 
In this segment, we will review case studies on data mining in African communities. These projects appear beneficial on the surface level, however, they embody a colonial nature that is deeply embedded in our world structures. Before we review the case studies, we will review what data mining is in this context.

\subsection{Defining Data Mining}

What is the difference between data sharing and data mining? Data sharing implies that there is an ownership of the data and an openness or agreement to share information. Data mining gives the impression of taking without asking, with no acknowledgement or compensation, while the miners of the data are the sole beneficiaries. 

Data colonialism is closely tied to data mining, an enactment of neo-colonialism in the digital world which uses data as a means of power and manipulation. Manipulation runs rampant in this age of misinformation, which we have seen heavily at play in recent times as well as throughout history, usually by playing on emotions and fear-mongering to influence public opinion. 

\subsection{ Data Mining in the African Context}

Data mining is a prime example of conflicting principles of AI ethics. On one hand, it is the epitome of transparency and a crucial element to scientific and economic growth. On the other hand, it brings up serious concerns about privacy, intellectual property rights, organisational and structural challenges, cultural and social contexts, unjust historical pasts, and potential harms to marginalised communities.\citep{Abebe2021NarrativesAfrica} 

The term data colonialism can be used to describe some of the challenges of data sharing, or data mining, which reflect the historical and present-day colonial practices such as in the African and Indigenious context.\citep{Couldry2019DataSubject} When we use terms such as ‘mining’ to discuss how data is collected from people, the question remains: Who benefits from this data collection? 

The aggregation and use of data can paradoxically be harmful to communities from which it is collected. Establishing trust is challenging due to the historical actions taken by data collectors while mining data from Indigenious populations. What barriers exist that prevent data from being of benefit to African and Indigeneous people? We must address the entrenched legacies of power disparities concerning what challenges they present for modern data sharing.\cite{Abebe2021NarrativesAfrica} As of 2021, the Open Government Partnership (OGP) lists fourteen members and ten states which have enacted data protection in Africa: Burkina Faso, Cabo Verde, Côte d'Ivoire, Ghana, Kenya, Liberia, Malawi, Morocco, Nigeria, Senegal, Seychelles, Sierra Leone, South Africa, and Tunisia.\citep{OPG2021DataProgress} The OPG is a charter which aims to liberate government-controlled data and focuses on the principles of transparency, accountability and participation.\citep{OPG2021DataProgress} 

\subsection{Data Mining Case Study}

One problematic example of data mining is non-government organisations (NGOs) that try to ‘fix’ problems for marginalised ethnic groups, but can end up causing more harm than good. For instance, a European NGO had attempted to address the problem of access to potable water in Burundi, while testing new water accessibility technology and online monitoring of resources.\citep{Abebe2021NarrativesAfrica}

The NGO failed to understand the perspective of the community on the true central issues and potential harms. Sharing the data publicly, including geographic locations, put the community at risk, as collective privacy was violated and trust was lost. Note that Burundi is not on the above list, and does not yet have data protection laws in place, nor a definition of personal data under Burundi law, however companies are required to gain consent prior to transferring personal data to third parties under some sectoral laws.\citep{OPG2021DataProgress} In the West we often think of privacy as a personal concern, however collective identity serves as a great importance to a multitude of African and Indigenious communities.\citep{Abebe2021NarrativesAfrica} 

Another case study in Zambia observed that up to 90\% of health research funding comes from external funders, meaning the bargaining power gives little room for Zamban scholars. In the study, power imbalances were reported in everything from funding to agenda setting, data collection, analysis, interpretation, and reporting of results.\citep{Vachnadze2021ReinforcementKaas} This example further exhibits the understanding that trust cannot be formed on the foundation of such imbalances of power. 

Many of these research projects lead with good intentions, yet due to a lack of forethought into the ethical use of data, during and after the project, unforeseen and irreparable harm can be done to the wellbeing of communities. This creates a hostile environment upon which  to build relationships of respect and trust.\citep{Abebe2021NarrativesAfrica} It exemplifies neo-colonialism in action: systems of victimisation and dependence which ultimately cause harm to the people it proposes to help. 

To conclude this case study, we can pose the ethical question: Is data sharing actually beneficial? First and foremost, local communities must be the primary beneficiaries of responsible data sharing practices. It is important to have specificity and transparency around who benefits from data sharing, and to make sure that it is not doing any further harm to the people behind the data.
Heretofore, neo-colonialism, ID-SOV, global data law and data mining have been discussed. In the next section, the connection to how we steer global data governance towards protecting the natural environment will be explored, examining case studies on metal mining in Mexico and exploitation of resources in Colombia. First, we will review the current status of the environment and climate change, including how data and law are used to help Indigenous communities in contemporary times.

\section{Environmental Concerns and Indigenous Data Sovereignty}

What is the connection between data governance, the environment, and the rights of Indigenous peoples?
In the intersection between rising technologies based on data and the need to protect the earth and all its peoples, we turn to Indigenous Data Sovereignty (ID-SOV). Indigenous communities are disproportionately affected by extraction and exploitation practices throughout the world, which harms the individual and the Earth.\citep{Portner2022ClimateVulnerability} How can ID-SOV help?
Before colonisation, Indigenous peoples had sovereignty over their data, in forms such as art and storytelling, and one thing that links the diverse populations of Indigenous communities is their inherent connection to the natural world. These connections to data and the Earth have been corrupted by colonisation, which has historically legitimised and legalised the exploitation of land, resources, and the people themselves.

\subsection{Data Mining and Mineral Mining}

We highlight the example of mining metals from the earth to reflect the term data mining, as on the surface, mining appears to be a part of our modern world and how we extract metals necessary for many things that we use daily, such as smartphones and computers; however, when we look closer, we can see the complexity and the harms that come to communities and the natural environment in this extraction process. Data mining is different from metal mining in several ways; one being that data is not something that occurs naturally, but that must be produced, by individual people. Data mining is often compared to extracting resources, or as a sort of modern-day land grab.\citep{Couldry2019DataSubject} Metal mining also has negative implications for Indigenous data protection and protection of Indigenous peoples’ health and environment, a further implication of the colonial nature of the practice. Prioritising ID-SOV would help to mitigate harm in these areas.

\subsection{Climate Change, Data, and Indigenous Rights}

The world is aware that we are in a climate crisis, and yet we have the technology and means to solve it. Microsoft Research published an article titled “Tackling Climate Change with Machine Learning” which addresses suggestions for switching to low-carbon electricity sources, thus lowering carbon emissions, removing CO2 from the atmosphere, and generally incorporating ML practices to improve standards for climate impact in the areas of Education, Transportation, Finance, Agriculture, and Manufacturing.\citep{Rolnick2019TacklingLearning} A recent UN report\citep{Portner2022ClimateVulnerability} on climate change provides the data and research needed to turn the ship around if only policymakers and corporate powers can get on board and make serious changes, and soon. The report based its findings on scientific knowledge as well as Indigenous and local knowledge to reduce risks from human-induced climate change while evaluating climate adaptation processes and actions.\citep{Portner2022ClimateVulnerability} 

The path forward includes a lot of moving parts, and it is vital to highlight the importance of Indigenous rights in this process. Indigenous ideologies are frequently based on the core tenet that there is really no separation between people and the Earth. On the other hand, governments and laws globally have a long colonial history of treating land as a commodity to be exploited for profit, endorsed by development discourses.\citep{Rojas-Paez2021NarrativesSetting}

It is necessary to stress the importance of mitigating undue harm to Indigenous peoples when applying data and machine learning technologies to help the planet. Sometimes when we set out to do good, it ends up hurting people unintentionally, and inadvertently reproducing colonial constructs. It is a delicate balance when outside researchers approach issues where they think they know the answers and what is best for others, but if we collaborate efforts and take responsibility for our shared planet, spending more time listening and less prescribing, ultimately, cooperation will be what helps all people the most. 

\subsection{Highlighting The Native American Rights Fund
}

The Native American Rights Fund (NARF)\citep{NativeAmericanRightsFund2022ProtectResources} stands to protect the rights of indigenous populations in the US, and their current environmental work primarily concerns climate change. On the international level, NARF has represented the National Tribal Environmental Council and the National Congress of American Indians (NCIA) via the United Nations Framework Convention on Climate Change, ensuring the protection of indigenous rights in international treaties and agreements governing greenhouse gas emissions. NARF represents Indigenous peoples in court cases, such as in a case representing Alaskan Native communities against energy companies for damages. They also help Indigenous peoples relocate when necessary, as the impact of climate change in Alaska has escalated to become immense and immediate.\citep{NativeAmericanRightsFund2022ProtectResources} NARF uses the law to help Indigenous peoples in America, standing against those who benefit from exploiting their lands and resources.

Environmental problems including climate change, habitat destruction, mining wastes, air pollution, hazardous waste disposal, illegal dumping, and surface/groundwater contamination cause an array of health and welfare risks to Indigenous peoples. NARF is one organisation helping Indigenous peoples to protect the environment as a top priority and helps to enforce laws such as the Safe Drinking Water Act, the Clean Air Act, and the Clean Water Act.\citep{NativeAmericanRightsFund2022ProtectResources} 

NARF has not yet helped to enforce laws around data protection and data mining practices for Indigenous people, however, they are in a position to assist on a legal level. These laws are nascent in their development and governments globally are only beginning to find consensus on their implementation. NARF is well-positioned to lead the way or join the movement that is beginning to ensure Indigenous data, history, land, and representation is protected in the digital future.

\subsection{Indigenous Policy and Case Studies in Mexico and Colombia}

Indigenous Data Sovereignty (IDS) is not only relevant but necessary for creating fairer governance and a more prosperous future for Indigenous peoples,\citep{Rodriguez2021IndigenousPerspectives} as was shown repeatedly throughout the book, Indigenous Data Sovereignty and Policy.\citep{Walter2021INDIGENOUSPOLICY} We will rely on two case studies from the book to exemplify this further and connect them to environmental concerns. We will also review recent laws that serve to protect Indigenous rights and cultural heritage in Colombia and Mexico to show how their governments have been taking action in these areas.

The chapter by Oscar Luis Figueroa Rodriguez focused on IDS in Mexico and examined unsolved priorities from history that involved the use of data - particularly the way that information and knowledge have been generated, transformed, controlled, and exploited across many contexts, indicating a need for ID-SOV.\citep{Rojas-Paez2021NarrativesSetting} Another chapter of the book highlighted Colombia’s struggle with ID-SOV,\citep{Rojas-Paez2021NarrativesSetting}  and introduced a 2019 ruling (the Jurisdicción Especial para la Paz “Special Jurisdiction for Peace” or JEP) which declared nature to be a victim of Colombia’s conflict, pointing to affected ecosystems including rivers and the need for them to gain legal protection.\citep{Rojas-Paez2021NarrativesSetting} 

Access to resources such as water is globally justified to be controlled selectively and exclusively, as nature is increasingly commodified. However, Indigenous narratives stand in opposition to this commodification.\citep{Johnson2016TheCriminology} \citep{Rojas-Paez2021NarrativesSetting} By legally protecting rivers and other ecosystems, historically normalised exploitation practices must come to an end. A century ago, in the 1920s, the implementation of large-scale economic projects led to the legitimization of direct violence against Indigenous peoples, for example, in the Amazon and northern region of Santander, oil and rubber plantations resulted in several Indigenous communities’ disappearance, through enslavement and assassination.\citep{Rojas-Paez2021NarrativesSetting} 

This was not a practice of Indigenous data mining but of erasure. Not only were Indigenous peoples and their knowledge misrepresented, but they were being wiped out. The only interest at the time was on what resources could be extracted from the land and how much it was worth. As there is now more of an interest in collecting Indigenous data as another form of resource mining, ID-SOV holds great importance in regard to Indigenous communities for it stands to mitigate “. . . demands for territorial rights, food sovereignty and access to natural resources such as water.”\citep{Rojas-Paez2021NarrativesSetting} 

Indigenous worldviews which hold that humans are a part of the land and cannot be separated from it have been undermined by the ideology that land is a commodity to be exploited for economic purposes. Rojas-Páez and O’Brien bring up the question, “. . . why is the human cost of the expansion of the extractive economy not challenged in countries whose Indigenous communities are still facing extermination?”\citep{Rojas-Paez2021NarrativesSetting} The authors turn to scholar Julia Suarez-Krabbe, who commented on the invisibility of the impact of colonial practices in places like Colombia and explained that “. . . the force of colonial discourse lies in how it succeeds in concealing how it establishes and naturalises ontological and epistemological perspectives and political practices that work to protect its power.”\citep{Rojas-Paez2021NarrativesSetting} \citep{Suarez-Krabbe2016RaceSouth}  However, recent rulings like the JEP work to recognize Indigenous ontologies, how their data is represented, and to protect the land.

\subsubsection{Data Protection and Indigenous Rights in Colombia}

“The protection of personal data is a constitutional and fundamental right in Colombia,” stated Carolina Pardo, partner in the corporate department of Baker McKenzie in Colombia.\citep{Pardo2021ColombiaOverview} Her article, Colombia Data Protection Overview in DataGuidance, references the Congress of Colombia enacted Statutory Law 1581 of 2012, which Issues General Provisions for the Protection of Personal Data (‘the Data Protection Law’), which ‘develop the constitutional right of all persons to know, update, and rectify information that has been collected on them in databases or files, and other rights, liberties, and constitutional rights referred to in Article 15 of the Political Constitution.’\citep{Pardo2021ColombiaOverview}

Colombian Indigenous inspectors have been named to monitor natural resources on reservations since 1987. In 1991, Colombia approved their new Constitution recognizing Indigenous rights, including ethnic and cultural diversity, languages, communal lands, archeological treasures, parks and reservations which they have traditionally occupied; and measures to adopt programs to manage, preserve, replace and exploit their own natural resources.\citep{UniversityofMinnesotaHumanRightsCenter1995UniversityLibrary} 

The Colombian government’s efforts and commitments to strengthen the dialogue on human rights have been recognized by political figures of the European Union. Patricia Llombart, Colombia’s EU Ambassador, stated that Colombia has shared values with the EU and is seen as a reliable and stable partner. Where the EU has been involved, international agreements which include protecting Indigenous rights as well as labour rights and rights for children have been signed in Andean countries.\citep{BlancoGaitan2019ChallengesProtection}

\subsubsection{Mexico: the example of Mineral Mining and ID-SOV}

Turning now to the Mexican chapter, we see a similar history echoed. Mineral mining is one extractive process which profoundly impacts Indigenous communities and has only been promoted by recent presidents in Mexico. In the last 12 years, 7\% of Indigenous territories have been lost for the sake of mining alone, frequently without even informing the Indigenous communities.\citep{Valladares2018ElIztapalapa} \citep{Rodriguez2021IndigenousPerspectives}  Without the knowledge or permission of local Indigenous peoples, external actors have historically conducted research to better understand the values of natural resources in Indigenous territories, demonstrating a lack of understanding of the implications of exploiting things such as minerals, timber, wildlife, plants, and water for the people who live there, in terms of health and environmental consequences, infrastructure, and investments.\citep{Rodriguez2021IndigenousPerspectives} However, we also see how the government has stepped up in recent times to protect Indigenous rights to their own culture and data, as well as the rights of Afro-Mexican peoples.

Mining metals from the earth is necessary for much of the goods and technologies we know and love today, however, there is a price to pay. The major cost falls on the people who live where these metals are extracted, or, where they used to live if they had their territories taken away for the purpose of mining.\citep{Rodriguez2021IndigenousPerspectives} 

We use mining as an example, which clearly shows the need for IDS and the consequences, as communities were neither considered nor informed about the extremely invasive methods and exploitation techniques involved in metal mining from their land in Mexico, including not only the use of heavy machinery but massive lixiviation processes mainly with sodium cyanide, which several European countries have forbidden.\citep{Boege2013ElXXI} \citep{Rodriguez2021IndigenousPerspectives} However, new laws have recently been enacted to protect Indigenous communities and Afro-Mexican communities and their heritage. January of this year (2022) saw a vote by the Mexican Congress to approve the Federal Law for the Protection of the Cultural Heritage of Indigenous and Afro-Mexican Peoples and Communities.\citep{Hermosillo2022MEXICO:Communities} This law includes protecting Indigenous and Afro-Mexican communities and their rights to property and collective intellectual property, traditional knowledge and cultural expressions, including cultural heritage, in an “. . . attempt to harmonise national legislation with international legal instruments on the matter, trying to give a seal of ‘inclusivity’ to minorities.”\citep{Schmidt2021NewMexico}

Cultural heritage relates directly to data about communities, which is where collective sovereignty over data comes into play. “Intangible cultural heritage is defined as the uses, representations, expressions, knowledge, and techniques; together with the instruments, objects, artefacts, and cultural spaces that are inherent to them; recognized by communities, groups, and, in some cases, individuals as an integral part of their cultural heritage.”\citep{Schmidt2021NewMexico} New and relevant definitions such as “cultural heritage” are helpful to guide third parties on identifying whether authorization is necessary for use of Indigenous or Afro-Mexican cultural heritage, as failure could result in infringements and/or felonies under Mexican law.\citep{Hermosillo2022MEXICO:Communities} This could help with the risks of data mining for these communities. We can see how the representation of cultural heritage in Mexico as well as Colombia has gained importance and legal protection, which is vital to the conversation about data. Cultural heritage represents data about a collective. By protecting cultural heritage, the lands and natural resources of Indigenous communities are also protected. 

Indigenous Data Sovereignty has a place as the laws change to protect Indigenous rights. ID-SOV could help “. . . fill the gap regarding the lack of evaluations as an appropriate approach in the design and implementation of monitoring, evaluation, and learning (MEL) local systems, controlled and used by Indigenous communities.”\citep{Rodriguez2021IndigenousPerspectives} 

Rodriguez went on to list recommendations from the Organisation for Economic Co-operation and Development (OECD) to move forward on these issues.

The OECD recommends four main areas to strengthen Indigenous economies:
\begin{itemize}
	\item  improving Indigenous statistics and data governance
	\item  creating an enabling environment for Indigenous entrepreneurship and small business development at regional and local levels
	\item  improving the Indigenous land tenure system to facilitate opportunities for economic development
    \item  adapting policies and governance to implement a place-based approach to economic development that improves policy coherence and empowers Indigenous communities \citep{OECD2019LinkingDevelopment}\cite{Rodriguez2021IndigenousPerspectives}
\end{itemize}

Lists like this are helpful, however, they must be approached with care and in communication with the people which they aim to help. These steps must be implemented by Indigenous peoples themselves with the support of organisations such as the OECD.

Through exploring these case studies from Mexico and Colombia, it is clear that ID-SOV could fill the gaps in considering public policies for data governance for Indigenous peoples, and that there is a need to remediate three main data challenges: data collection, data access, and relevance in order to access, use and control their own data and information.\citep{Rodriguez2021IndigenousPerspectives} This is something that must be understood for data governance around the world, and to note that there are different local concerns in different regions, although all have been negatively influenced and impacted by long-standing exploitative colonial practices. It is imperative that we continue to educate ourselves and question broader narratives that stem from colonial roots.

In the final section, it is time to address how to understand the role of data governance and decolonisation for the Global South, and complexify what the Global South means. We will also visit our final case studies, which, possibly ironically, address decolonising research with Inuit communities in the far North. 

\section{Global South and Counter Powers}

Throughout this paper, we have visited case studies in Africa, Colombia, and Mexico, however we have yet to directly address how we situate the Global South. In this section, we learn that there are multiple Souths, and not all of them fall below the equator. The case studies on the Inuit discuss decolonisation of data where it starts and where it ends. It starts with the research which data collectors do, and it ends with what it is like to live in the digital world for people who traditionally live in accord with nature, but whose knowledge is also based around adaptation. How can we all learn to adapt in this rapidly changing world?

\subsection{The Decolonial Turn and Pluralising the Souths }
In Nick Couldry and Ulises A. Mejias’s most recent article, the decolonial turn was discussed in detail, which links data extraction everywhere, whether in the Global North or Global South, to the colonial underpinnings of capitalism by looking through  “. . . the long-term historical lens of attempts to justify the unequal distribution of the world’s resources that began in earnest 500 years ago”.\citep{Couldry2021TheHeading} This, they said, is crucial to acknowledge when addressing contemporary discourses such as Big Data and AI for Social Good, which need to account for the implications of governing human life and freedom through data extraction practices which are colonial in nature.\citep{Couldry2021TheHeading} 

Stefania Milan and Emiliano Treré complicate the idea of the Global South by pluralizing it to be the Souths, as tracing disempowerment geographically through data does not always follow a North-South dichotomy. They make the case that there are undeniable inequalities seen where datification is weaponized by institutions and corporations to manage people, hitting harder where human rights and laws are the most fragile.\citep{Milan2019BigUniversalism} 

The plurality of the Souths extends into the North and West, for discrimination and inequality know no boundaries, and people who are in any way different or silenced can be found everywhere. That is why in our discussion on the Global South, we will actually be reviewing case studies from the far North - specifically among the Inuit in Northern Canada - to shed light on counter powers in the form of decolonising research and understanding the impact of datification and digitisation on Indigenous communities. 

\subsection{Decolonising Research to Decolonise Data: Digital Storytelling
}

Decolonising data starts with decolonising research, the process of collecting data. Digital storytelling is a process involving immersive workshops where the relationship of researcher-researched transforms into teller-listener, and personal stories and narratives are related with a mix of voice, video, photographs, artwork and music to create a sort of first-person mini-movie.\citep{CunsoloWillox2013StorytellingWisdom} 

In the article by Ashlee Cunsolo Willox and others, a mix of Indigenous and non-Indigenous individuals teamed up in 2009 in Northern Canada to engage with a remote community and develop a digital narrative method which examined the connection between climate change and health and well-being by uniting digital media with storytelling as a way to celebrate the individual and the collective.\citep{CunsoloWillox2013StorytellingWisdom}  The authors found that “. . .by uniting the finished stories together, a rich, detailed, and nuanced tapestry of voices emerge providing context and depth to localised narratives and collective experiences.”\citep{CunsoloWillox2013StorytellingWisdom} 

Digital storytelling requires a high level of trust, and has great potential as a more participatory and democratic form of social research, however it brings up a lot of personal and political questions while creating this raw form of narrative “data.” Researchers still have to decide which stories to share, and monitor their interference with how the community is represented to give voice to peoples and issues which are generally silenced, without perpetuating stereotypes or misunderstandings. Digital storytelling can disrupt, alter and/or reverse power dynamics in narrative research by removing the researcher as the teller of others’ stories and build a powerful source of data coming straight from lived experiences of individuals, creating “. . . the opening to listen, reflect, learn, trust, and then listen again.”\citep{CunsoloWillox2013StorytellingWisdom} 

Decolonising the way that research is carried out is a crucial step in decolonising data practices, as the collection of data is where data colonisation starts. Narratives can get convoluted, if not completely misrepresented. It is vital to build relationships based on mutual trust, respect, and communication to ensure that people are being represented correctly, that they are informed on things that affect them, and that they are the primary beneficiaries of their own data. 

\subsection{ Digital Colonialism and Inuit Knowledge
}
Finally, this paper explores a study about the traditional knowledge of the Inuit in order to understand how digital colonialism can affect communities. Inuit traditional knowledge is referred to as IQ (Inuit Qaujimajatuqangit) which represents a set of skills to be constantly practised and adapted to a changing world. As opposed to knowledge that can be held or recited, rather, IQ is cultivated by experiences in the natural world and from elders, as unlike Western forms of objective knowledge, there is not the separation between knowledge and knowledge-holder,\citep{Laugrand2010InuitCentury} and it cannot be learned from reading or watching videos.\citep{Young2019TheColonialism} In Jason C. Young's research in Igloolik, he heard repeatedly in interviews the sentiment that the internet is killing IQ culture, due to community members spending less time on the land and with elders and more time online.\citep{Young2019TheColonialism}

“Digital engagement is undermining two key aspects of the IQ system—embodied socialisation and experiential learning out on the land. Time spent online can trade off with embodied play outdoors, visits at the homes of other Iglulingmiut, and visits to elders outside of the community.”\citep{Young2019TheColonialism} 

Due to the nature of IQ being about adapting to change, community members seek ways to balance technology and lived experiences. One example is a Facebook group called Nunavut Hunting Stories of the Day, which allows Nunavummiut to share hunting stories and knowledge meant to inspire others, mainly youth, to get out on the land themselves and have their own learning experiences.\citep{Young2019TheColonialism}  Efforts like this application should be promoted. Additionally, encouraging the importance of community leaders and elders in conversations about the protection of rights, the environment, and data need to be prioritised. 

There are ways that data, AI, and related technologies can be used for promoting culture and data sovereignty, however it does not come automatically. What comes automatically are the patterns laid forth by colonialism, capitalism, consumerism, and development. When considering ID-SOV as a core tenet for designing global data governance, it is important to see it from all sides. Data should be shared back to the communities to whom it belongs, and it should benefit them, not harm them. The ultimate goal should be to bring people together, not to push them further apart and further from the land on which they depend. Data should be a compliment to life, not what governs life itself. 
\section{Conclusion}

“For the master’s tools will never dismantle the master’s house. They may allow us to temporarily beat him at his own game, but they will never enable us to bring about genuine change.”
― Audre Lorde 1979\citep{Lorde2018TheHouse}

In the end, we will face innumerable challenges decolonising anything if we are still building within systems of development and capitalism, as these are all the ‘master’s tools.’ New tools will be necessary to build in the cracks and develop global data governance that works for everyone without leaving vulnerable people at the margins. Colonialism is internalised, for all of us, and we are all colonised. It is one thing that connects us. We know where we each sit within this system, and no matter where that is, we may point to it and declare that it is wrong, meanwhile continuing to play our respective parts. Decolonising is going to take a lot of work, patience, and care, in addition to a lot of understanding of the complex history that we come from, in order to steer data governance toward a brighter future that protects all people and the Earth as we face the plethora of social and environmental issues in front of us. 

This paper has given a brief overview of several related concepts that are important to consider for global data governance and decolonisation. Data colonialism and digital colonialism were reviewed, followed by Indigenous Data Sovereignty. Suggestions and potential solutions when considering global data law were explored, followed by the tensions related to data mining, which led us to drawing together environmental concerns with Indigenous rights and data laws. Finally, we complexified the Global South and further discussed decolonisation. Throughout, we highlighted important organisations and movements, such as NARF, the digital NAM, Technologies for Revolution, and the OECD to show which groups are poised to support ID-SOV going forward. We explored case studies in Africa, Colombia, Mexico, and Northern Canada to give a global perspective. Future research could dive much deeper into any of the areas discussed, and could be elaborated on for further understanding of these topics.

\label{sec:others}

\bibliographystyle{unsrtnat}
\bibliography{reference}  %%% Uncomment this line and comment out the ``thebibliography'' section below to use the external .bib file (using bibtex) .

\end{document}